\documentclass[]{spie}  %>>> use for US letter paper
\usepackage{amsmath,amsfonts,amssymb}
\usepackage{graphicx}
\usepackage{csquotes}
\usepackage[colorlinks=true, allcolors=blue]{hyperref}
\usepackage{lineno}
\title{AstroPix: Investigating the Potential of Silicon Pixel Sensors in the Future of Gamma-ray Astrophysics}

\author[a]{Isabella Brewer}
\author[b]{Regina Caputo}
\author[b,c]{Michela Negro}
\author[d]{Richard Leys}
\author[b]{Carolyn Kierans}
\author[d]{Ivan Peric}
\author[g]{Jessica Metcalfe}
\author[b]{Jeremy Perkins}
\affil[a]{University of Maryland, College Park, MD, USA}
\affil[b]{NASA Goddard Space Flight Center, Greenbelt, MD, USA}
\affil[c]{University of Maryland Baltimore County, 1000 Hilltop Cir, MD 21250, Baltimore, USA}
\affil[d]{ASIC and Detector Laboratory, Karlsruhe Institute of Technology, Germany}
\affil[g]{Argonne National Laboratory, Washington, DC, USA}
\authorinfo{Further author information: (Send correspondence to I. Brewer)\\I. Brewer: E-mail: ibrewer@umd.edu\\}

\begin{document} 
\maketitle

\begin{abstract}
This paper details preliminary photon measurements with the monolithic silicon detector ATLASPix, a pixel detector built and optimized for the CERN experiment ATLAS. The goal of this paper is to determine the promise of pixelated silicon in future space-based gamma-ray experiments. With this goal in mind, radioactive photon sources were used to determine the energy resolution and detector response of ATLASPix; these are novel measurements for ATLASPix, a detector built for a ground-based particle accelerator. As part of this project a new iteration of monolithic Si pixels, named AstroPix, have been created based on ATLASPix, and the eventual goal is to further optimize AstroPix for gamma-ray detection by constructing a prototype Compton telescope. The energy resolution of both the digital and analog output of ATLASPix is the focus of this paper, as it is a critical metric for Compton telescopes. It was found that with the analog output of the detector, the energy resolution of a single pixel was 7.69 $\pm$ 0.13\% at 5.89 keV and 7.27 $\pm$ 1.18\% at 30.1 keV, which exceeds the conservative baseline requirements of 10\% resolution at 60 keV and is an encouraging start to an optimistic goal of $<$2\% resolution at 60 keV. The digital output of the entire detector consistently yielded energy resolutions that exceeded 100\% for different sources. The analog output of the monolithic silicon pixels indicates that this is a promising technology for future gamma-ray missions, while the analysis of the digital output points to the need for a redesign of future photon-sensitive monolithic silicon pixel detectors. 
\end{abstract}

\keywords{Silicon pixel detectors, Compton telescopes, high energy astrophysics}

\section{INTRODUCTION}
\label{sec:intro}  % \label{} allows reference to this section

AstroPix is a project investigating the potential of monolithic silicon (Si) pixels as a detector material in future space-based gamma-ray astrophysics. The beginning of this project, and the focus of this paper, involves testing existing monolithic Si pixel sensors with high-energy radioactive sources. Once the energy resolution and detector response of existing technology has been ascertained, the eventual goal of the AstroPix project is to create a small prototype Compton telescope with a monolithic-Si-pixel-based tracker. The exigence for this technology partly arises from a new generation of gamma-ray instruments on the horizon, particularly instruments focusing on the MeV band of 0.2 MeV to 10 GeV. This includes the All-sky Medium Energy Gamma-ray Observatory (AMEGO)\footnote{https://asd.gsfc.nasa.gov/amego/} project, a proposed mission answering the call of the 2020 Decadal Survey. AMEGO could thoroughly investigate events in the MeV range, an overlooked area on the energy spectrum, and provide support in the burgeoning field of multimessenger astrophysics\cite{mcenery}. Additionally, AstroPix-style detectors have the potential to assist with the detection of gamma-ray bursts (GRBs), also critical events for furthering multimessenger astrophysics. By virtue of the inherent properties of Compton telescopes, a Compton telescope such as AstroPix would be able to provide polarization data for GRBs\cite{Caputo1}.

Monolithic Si pixel sensors rely on Complementary Metal-Oxide-Semiconductor (CMOS), a common manufacturing technique used in commercial industry. Pixelated silicon sensors leverage High Voltage CMOS manufacturing processes to integrate charge collecting diodes, analog frontend and digital readout electronics in a shared substrate, in a so-called monolithic sensor architecture. This co-integration of detector and readout electronics is the main attraction of CMOS pixels. Double-sided silicon strip detectors (DSSDs), a technology analogous to Si pixels and a current candidate for future space-based gamma-ray missions such as AMEGO\cite{griffin}, are difficult to fabricate. Once made, DSSDs would have to be chained together, increasing the capacitance and therefore noise of the instrument. Si pixels, on the other hand, would reduce noise, which would improve energy resolution and jointly enhance other science parameters like angular resolution. Si pixels reduce the amount of passive material, improving reconstruction at low energies and reducing the financial costs of space-based missions by virtue of compactness. The potential advantages of CMOS technology are the motivation behind testing existing pixelated Si technology with photons. A customized monolithic pixel Si detector named ATLASPix had been previously developed for the ATLAS\cite{ATLASPix2020} collaboration at the European Organization for Nuclear Research (CERN)\footnote{https://home.cern}. We worked with our collaborators at CERN to evaluate ATLASPix's potential as a gamma-ray detector by exposing it to various radioactive photon sources. Collaborators at CERN had developed software to read out and control the detector, and provided measurements as a baseline.

\begin{figure}[!htbp]
    \centering
        \includegraphics[width=.4\textwidth]{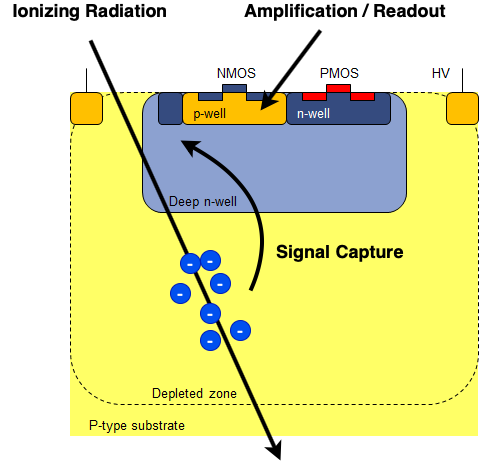}
        \caption{Diagram of an ATLASPix CMOS pixel. CMOS pixels co-integrate readout and detector capabilities, which allows for less passive material, lower power consumption, and less noise.}
        \label{setup_pic}
    \end{figure}
    
ATLASPix is currently optimized for the detection of Minimizing Ionizing Particles (MIPs)\cite{bethe1}\cite{bethe2}\cite{bloch} in a ground-based particle accelerator environment. The focus of this project is to ascertain how ATLASPix could be redesigned for the detection of high energy photons, specifically in the Compton regime as a new detector technology in a space-based telescope. With a precise measure of the position and energy deposits in a sequence of Compton scatters, Compton telescopes can reconstruct the incident photon's initial energy and direction; therefore, the energy and position resolution of the detector determine the quality of reconstruction, angular resolution, and sensitivity of a Compton telescope. Determining these design parameters is the first step in the larger AstroPix project. Eventually, a prototype Compton telescope will be constructed that must meet a few key requirements: it must achieve energy resolution $<$2\% at 60 keV, position resolution $\sim$250 $\mu$m, have a pixel size of 500 $\mu$m, and have a Si thickness of at least 500 $\mu$m. A new iteration of Si pixels optimized for high-energy photon detection and based on ATLASPix, AstroPix V1, have already been constructed and will soon undergo testing. This paper will focus on the testing of ATLASPix, the monolithic Si pixel detector optimized for MIPs. Energy resolution is of especial interest, but energy calibration and dynamic range will also be explored. The ATLASPix detector has two outputs, a digital and analog output, and the energy resolutions of these two outputs will be compared to see how future iterations of AstroPix can improve on detector design. This paper will detail the experimental set-up in section \ref{sec:setup}, detector response and energy resolution of ATLASPix in section \ref{sec:energy_calibration}, simulations of parameters for future versions of AstroPix in section \ref{sec:simulations}, and will then discuss results and the future of AstroPix in section \ref{sec:conclusion}.

\section{EXPERIMENTAL SETUP}
\label{sec:setup}

In order to determine the energy resolution and detector response of ATLASPix, the detector was illuminated by the sources listed in Table \ref{source_table}. Three sealed isotropic sources (Fe55, Cd109, and Ba133) and an x-ray source pointed at three different targets (Ge, Y, and Mo) were placed in front of the detector at different times. The x-ray source resulted in collimated beams.

\begin{table}[ht]
\caption{Radioactive sources used for testing, including their energies and approximate count rate.} 
\label{tab:fonts}
\begin{center}       
\begin{tabular}{|c|c|c|}
%% |l|l| to left justify each column entry
%% |c|c| to center each column entry
%% use of \rule[]{}{} below opens up each row
\hline
\rule[-1ex]{0pt}{3.5ex}  \textbf{Source} & \textbf{Energy (keV)} & \textbf{Counts per Second}  \\
\hline
\rule[-1ex]{0pt}{3.5ex}  Fe55 & 5.89 & 0.8 \\
\hline
\rule[-1ex]{0pt}{3.5ex}  Ge & 9.89 &  0.2 \\
\hline
\rule[-1ex]{0pt}{3.5ex}  Y & 14.96 & 0.2  \\
\hline
\rule[-1ex]{0pt}{3.5ex}  Mo &  17.5 & 0.2 \\
\hline
\rule[-1ex]{0pt}{3.5ex}  Cd109 &  21.99 & 1.0 \\
\hline
\rule[-1ex]{0pt}{3.5ex}  Ba133 &  30.97 & 0.6 \\
\hline
\end{tabular}
\end{center}
\label{source_table}
\end{table} 

The sources used for experimentation were chosen based on the dynamic range of ATLASPix. The dynamic range of the detector currently stands at around 5 keV on the lower end, as a result of the noise floor, and about 32 keV at the upper end, representing the highest energy source (Ba133) that we were able to see with the detector. The upper limit of the dynamic range is a result of the current thickness of the Si (100 $\mu$m), a problem that could be solved with thicker Si. AstroPix would ideally be sensitive to energy in the range of Compton scattering, 30 keV to 3 MeV. 

ATLASPix is comprised of four matrices, with each matrix made of 100 rows and 25 columns. Each pixel is 50 $\mu$m by 140 $\mu$m and 100 $\mu$m thick. The current experimental setup consists of the ATLASPix detector (the ATLASPixSimple version\cite{ivan_1}), a Control and Readout Inner tracking Board (CaRIBOu) data acquisition (DAQ) System, a Zynq ZC706 Field Programmable Gate Array (FPGA), and a custom adapter board \cite{Caputo1}. A DC voltage supply was used to power the FPGA and a high voltage supply set at 60V was used as the bias voltage for the detector. A jig to hold various radioactive sources was placed in front of the detector without blocking the source. To determine energy resolution, photon sources in the x-ray range were placed in front of the detector. 

\begin{figure}[!htbp]
    \centering
        \includegraphics[width=.49\textwidth]{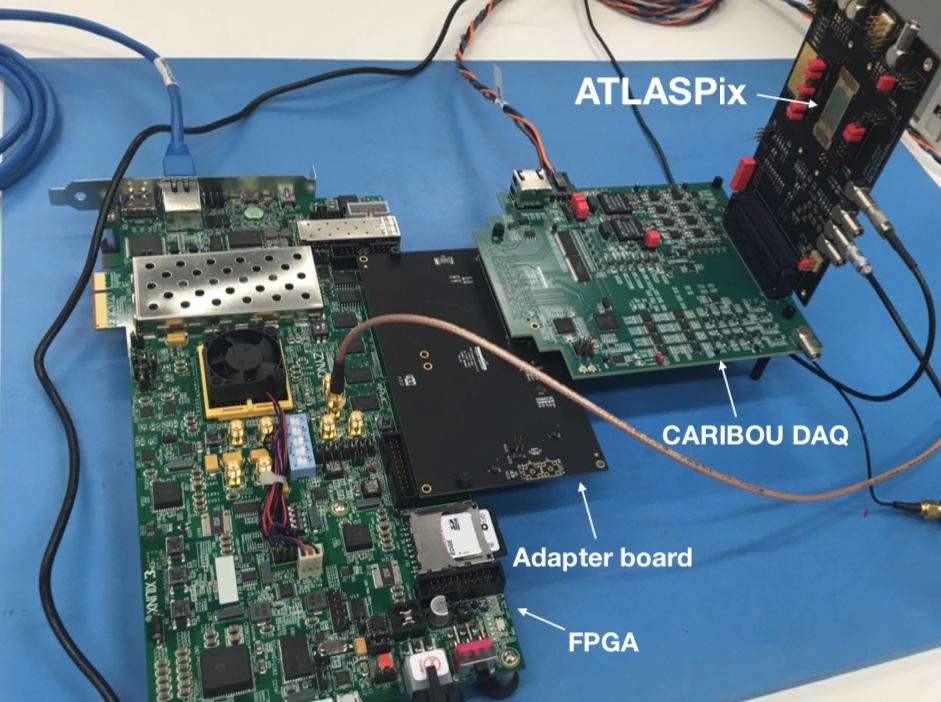}
        \caption{Experimental setup for ATLASPix.}
        \label{setup_pic}
    \end{figure}

There are two ways to read out the detector. The first is a digital output read out by the DAQ. The data comes in the form of a text file detailing hits from across all the pixels and relates a hit position, energy, and timestamp for each hit. The energy values from the DAQ file are a time over threshold (TOT) energy proxy; the TOT is capped at a six bit resolution. The ATLASPix digital output has a fine timing resolution of 25 ns that was optimized for a ground-based collider experiment. 

The other output for the detector is an analog voltage output which can be read out only from a single pixel at a time. During data collection, the voltage output was fed directly to an oscilloscope or multi-channel analyzer (MCA). Detector settings, such as pixel threshold, can be configured via command line software through the CaRIBOu DAQ system.

With each source, a data set was collected from the detector, either from the digital or analog output. The analog output was taken from individual pixels while the digital output was taken from the entire detector. In the case of the analog output, two different pixels, pixel (0,50) and pixel (12,50) (where pixel number is given by (x,y), see Figure \ref{heatmaps}), were used for measurements. The pixels were selected at random. More than one was used to examine difference in pixel response, but the amount of time required for tests limited analysis to two pixels.

\begin{figure}[!htbp]
\begin{minipage}[b]{0.45\linewidth}
\centering
\includegraphics[width=1.0\textwidth]{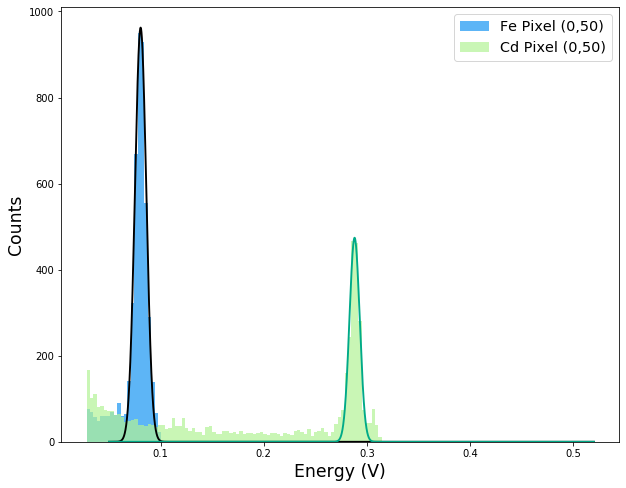}
\label{fig:figure1}
\end{minipage}
\hspace{0.5cm}
\begin{minipage}[b]{0.45\linewidth}
\centering
\includegraphics[width=1.0\textwidth]{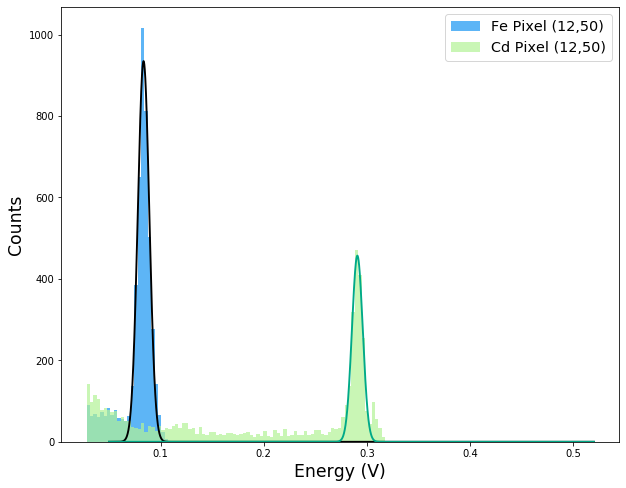}
\label{fig:figure2}
\end{minipage}
\caption{Histograms of the raw analog data (in volts) from two pixels. The same two sources, Fe55 and Cd109, were exposed to both pixels. A Gaussian was fit to each peak to compare peak positions between pixels. Left: Pixel (0,50). Right: Pixel (12,50).}
\label{compare_energy_res}
    \end{figure}

Plots showing histograms of the maximum value of each analog voltage output can be seen in Figure \ref{compare_energy_res}. Data was taken for two pixels exposed to the same two sources, Fe55 and Cd109. A Gaussian was fit to the data sets to determine how the central position $\mu$ and width of the peak $\sigma$ at identical energies changes between two pixels. Data was taken and fit separately. 

For pixel (0,50), the $\mu$ of Fe55 is 0.081 $\pm$ 0.005 V and for pixel (12,50), the $\mu$ of Fe55 is 0.084 $\pm$ 0.006 V. For pixel (0,50), the $\mu$ of Cd109 is 0.288 $\pm$ 0.005 V. and for pixel (12,50), the $\mu$ of Cd109 is 0.291 $\pm$ 0.005 V. Comparing the $\mu$s of the two sources, the pixels agree on $\mu$ placement within errors. This is encouraging, as it suggests consistent pixel response. More pixels need to be tested to verify a truly uniform pixel response, however, and it should be noted that while these pixels are from different columns, they inhabit the same matrix, and due to logic chaining, matrices may respond differently.

Analog data was then taken separately at every energy listed in Table \ref{source_table} for pixel (0,50). The centroid $\mu$ and sigma $\sigma$ of each fit of the raw data in volts was used to find the energy calibration for the analog output. A similar process was repeated for the digital output, although only four sources, Ge, Y, Mo, and Cd109, were used. Four sources were used due to the time constraints of collecting data and because of the low resolution of the digital output.

\section{Energy Response of ATLASPix}
\label{sec:energy_calibration}
\subsection{Energy Calibration}
\label{subsection_energy_scaling}

\begin{figure}[!htbp]
\begin{minipage}[b]{0.45\linewidth}
\centering
\includegraphics[width=1.0\textwidth]{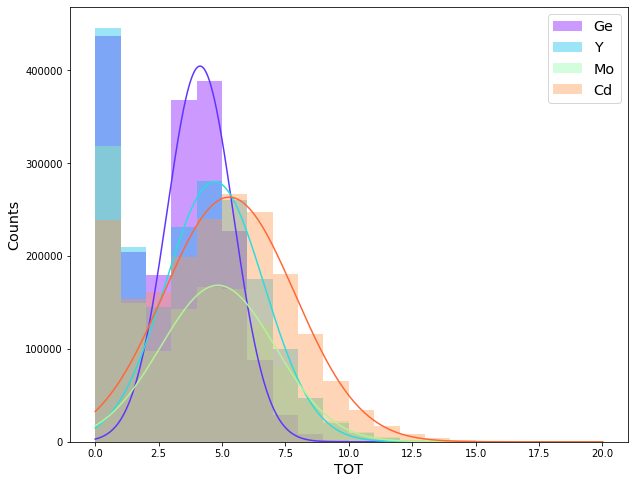}
\end{minipage}
\hspace{0.5cm}
\begin{minipage}[b]{0.45\linewidth}
\centering
\includegraphics[width=0.95\textwidth]{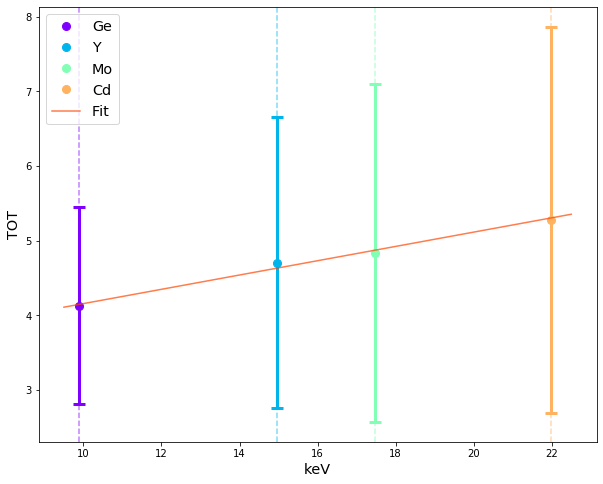}
\end{minipage}
\caption{Left: Distribution of the TOTs collected from all the pixels of the detector for different photon sources (Ge, Y, Mo, Cd109). Data was taken at separate times. Each histogram was fit separately with a Gaussian. Right: Energy calibration for TOT data with four sources. The best-fit Gaussian mean of the TOT distributions is compared with the theoretical keV values, and the error bars represent the best-fit Gaussian standard deviation of the same distributions.}
\label{digital_scaling_res}
    \end{figure}

The energy calibration details the relationship between the peak position $\mu$ found in volts or TOT (depending on whether the analog or digital output is being used) and the known energy line in keV. In Figure \ref{digital_scaling_res}, four sources were separately exposed to ATLASPix. The digital DAQ file records the pixel number and TOT value from each hit; each TOT value is associated only with a single pixel. The TOT values at each source energy, which come from pixels across the entire detector, were plotted as histograms. A Gaussian was fit to the histograms to find respective peak location $\mu$ and width $\sigma$. Figure \ref{digital_scaling_res} (left) shows the histograms of the raw TOT data at each source energy fit with a Gaussian and (right) the relationship between the peak positions the the Gaussian fits and the theoretical keV values.

A noise peak occupies the first bin of Figure \ref{digital_scaling_res} (left). Each source used has a distinct energy line, and the sources Ge to Cd109 cover an energy range of about 10 keV to 22 keV; distinct peaks should be visible and separate, but the low resolution of the digital output results in a large width $\sigma$ for each fit, making the energy lines almost indistinguishable. The energy resolution is defined from the full-width at half-maximum (FWHM) of the best-fit Gaussians, and it exceeds the 100\% for all four sources used for TOT measurements. The TOT measurements are from pixels across the entire detector, so some broadening of the peak may be due to disparity in pixel response. 
    
The large error bars on the left in Figure \ref{digital_scaling_res} represent the sigma found from fitting the TOT histograms and is the result of the six bit resolution of the digital output. Due to the uncertainties associated with each data point, a first order equation was sufficient to fit the digital energy calibration. In addition to examining detector response with the digital output, detector response with the higher resolution analog output was quantified.

Pixel (0,50) was illuminated by all the sources listed in Table \ref{source_table}. The analog voltage output of the pixel was recorded about 5000 times for each source. For each source, the maximum values in volts were then histogrammed and the peak position $\mu$  and was obtained from a Gaussian fit. 

\begin{figure}[!htbp]
    \centering
        \includegraphics[width=.48\textwidth]{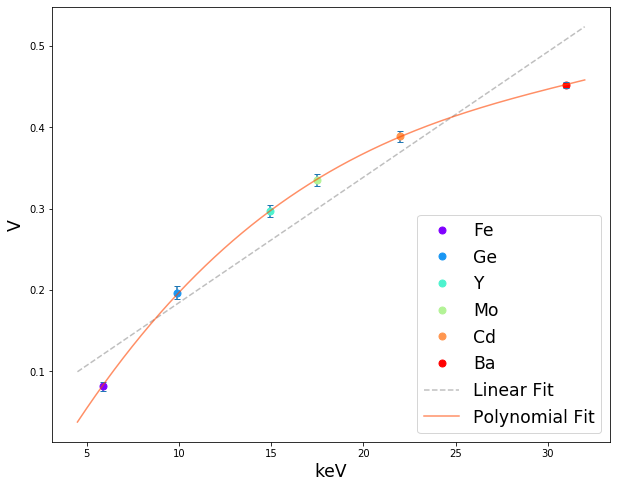}
        \caption{Energy calibration for pixel (0,50), with both a linear and three degree polynomial fit shown. A histogram of the raw analog voltage data was plotted for each source; a Gaussian was then fit to each histogram. The peak positions of these histograms in volts were compared to the theoretical energy values in kev to find the energy calibration. A three degree polynomial was found to more accurately represent the detector response than a linear fit. The error bars represent the sigma of the Gaussian fit to the raw data.}
        \label{analog_scaling}
    \end{figure}

In Figure \ref{analog_scaling}, the peak position in volts is compared to the theoretical keV value at each energy. Different fits of the volts to keV relationship were compared to determine the detector response. Although a two degree polynomial fit had a reduced $\chi^2$ of 1.1 and was found statistically to be sufficient for our data, a three degree polynomial fit allowed us to reconstruct energies within 1\% of the theoretical value (compared to the ~5\% mismatch obtained assuming a second-order polynomial relation). The three degree polynomial fit was therefore used for energy calibration. A linear detector response is optimal, but it is unsurprising that the energy calibration is non-linear. The non-linear detector response is due to readout electronics or perhaps detector material. 

\subsection{Energy Resolution}

A figure of merit for gamma-ray science is energy resolution, especially for Compton telescopes, as it is essential for event reconstruction. To measure the analog energy resolution, each source in Table \ref{source_table} was exposed to the detector. For each source, 5000 voltage outputs were recorded from a single pixel and the maximum value of each output was found. A noise sample, or the average of the maximum output when no source is present, was subtracted off each measurement. The voltage output distributions can be translated into energy distributions by applying the relation found through the calibration study described in Section \ref{subsection_energy_scaling} (see Figure \ref{analog_scaling}). A Gaussian was fit to the scaled histogram at each source energy, with the FWHM from the fit used to determine energy resolution. All analog data was fit separately. The analog output of a single pixel (0,50) was used for energy calibration because each pixel has a slightly different response.

\begin{figure}[!htbp]
    \centering
        \includegraphics[width=.49\textwidth]{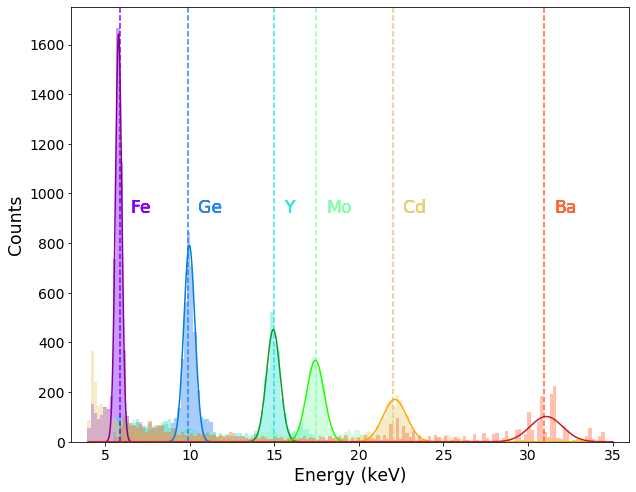}
        \caption{The energy resolution of pixel (0,50) at six different energies. All data was collected, at different times, from the analog output. In order to scale the data, the energy calibration from Figure \ref{analog_scaling} was applied to each data set. The energy resolution was derived from the full-width half-max (FWHM) of a Gaussian fit from each scaled data set. The lower resolution of the Ba133 data is due to resolution restraints from the oscilloscope used, not from the intrinsic resolution of the detector.}
        \label{analog_resolution}
    \end{figure}

With scaling applied, the energy resolution of a single pixel (0,50), plotted in Figure \ref{analog_resolution}, was found to be 7.69 $\pm$ 0.13\% at 5.89 keV and 7.27 $\pm$ 1.18\% at 30.1 keV. This exceeds our conservative goal of having an energy resolution of less than 10\% at 60 keV for AstroPix\cite{Caputo1}, and is a promising start to our goal of an eventual energy resolution of $<$2\% at 60 keV. Figure 6 shows the energy resolution of pixel (0,50) for the different tested energies. All the resolutions within errors exceed the minimum energy resolution requirement of 10\%. The error of each resolution was derived using error propagation from the peak width and peak position found in Figure \ref{analog_resolution}. That the analog output already meets energy resolution requirements is encouraging regarding the future of CMOS detectors in gamma-ray astrophysics. 

\begin{figure}[!htbp]
    \centering
        \includegraphics[width=.49\textwidth]{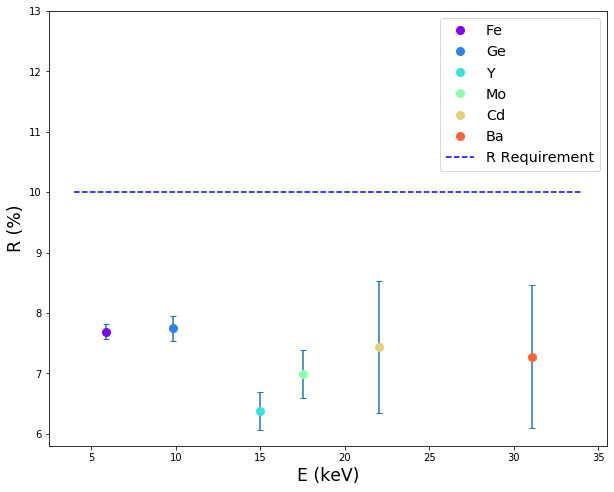}
        \caption{The resolution of pixel (0,50) plotted at each source energy. The horizontal dashed line represents the minimum resolution requirements of AstroPix, and the error bars are derived using error propagation from the width $\sigma$ and peak position $\mu$ found from the Gaussians seen in Figure \ref{analog_resolution}.}
        \label{e_v_r}
    \end{figure}

The energy resolution from the digital output of the detector is less auspicious, and highlights an opportunity for future iterations of AstroPix to improve in this area. TOT measurements from all pixels, seen in Figure \ref{digital_scaling_res}, resulted in energy resolutions of more than 100\% at each energy. Future versions of AstroPix would be able to improve digital energy resolution; AstroPix, for example, doesn't need the high resolution timing that ATLASPix currently possesses, and cutting back on this functionality would leave room for improved energy resolution. Given that energy resolution is a critical metric for Compton telescopes, the combined results of the analog output resolution-- which reveals that a desirable energy resolution is attainable-- and the digital output-- which reveals that certain aspects of the detector are not yet optimized for this type of data collection-- only increase the exigence of AstroPix.

\subsection{Hit Position and Clustering}

\begin{figure}[!htbp]
\begin{minipage}[b]{0.45\linewidth}
\centering
\includegraphics[width=0.5\textwidth]{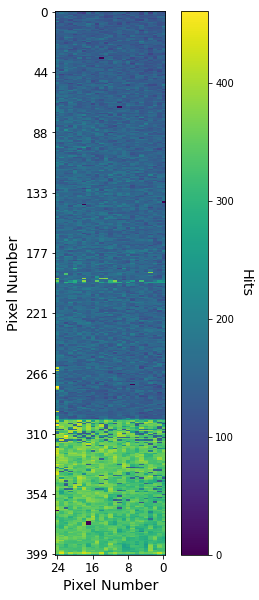}
\label{fig:figure1}
\end{minipage}
\hspace{0.5cm}
\begin{minipage}[b]{0.45\linewidth}
\centering
\includegraphics[width=0.51\textwidth]{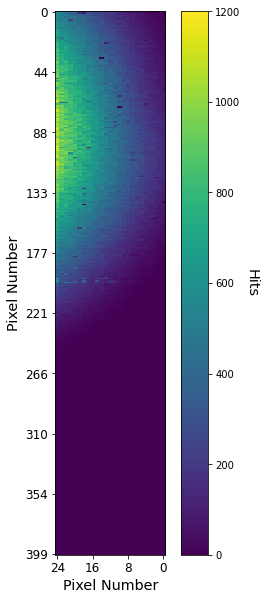}
\label{fig:figure2}
\end{minipage}
\caption{Heatmaps of the ATLASPix detector showing the distribution of hits when the detector is exposed to radioactive sources. Both axes represent pixel number in the x and y direction respectively. Data was extracted from the digital DAQ file, which records x and y position of each hit. In the heatmap from Cd109 (left), the matrix structure of the detector is visible, as the bottom matrix is noisier than the others. In the heatmap from Fe55 (right), the source can be more clearly seen, as the lower energy source resulted in a more concentrated pattern.}
\label{heatmaps}
    \end{figure}

Data was further examined to search for clusters of hits, which may be indicative of cross-talk or charge sharing. Data was extracted from the digital DAQ measurements to examine the number of clusters and cluster size. Heatmaps were also plotted to give a visual representation of how the detector responds to the sources. The heatmaps for Fe55 and Cd109 are depicted in Figure \ref{heatmaps}. 

A recursive search was used to find clusters from the DAQ files. Each hit in the DAQ file gives a hit position, timestamp, and TOT. For each hit, the text file was parsed to see if there were any hits in the eight pixels surrounding the initial pixel and within 25 \textit{ns}; if such a hit was found, the search was started again by designating the new hit the \enquote{pixel of interest} and searching the surrounding ring of 8 pixels of the new pixel (disregarding duplicates). 1000 hits were parsed from the Cd109 DAQ file, and 182 clusters were found, with an average cluster size of 2.75 pixels. Of the 1000 hits parsed from the Fe55 file, 11 clusters were detected, with an average cluster size of 2 pixels. Finding more clusters and larger cluster sizes with Cd109 was expected because it is a higher energy source than Fe55.

\section{Simulations}
\label{sec:simulations}

Simulations were carried out in the Medium Energy Gamma-ray Astronomy Library (MEGAlib)\cite{megalib} to ascertain ideal pixel size and thickness of the detector for a full-scale telescope based on the AstroPix design. %MEGALib simulations are key in optimizing how many layer of silicon the final revolutions of AstroPix should contain (Michela contributes this part?) 
We defined a baseline geometry of our instrument as composed of 50 layer tracker and ideal calorimeter to measure the energy and position of the Compton-scattered photon. 
This geometry is similar to the current AMEGO mission\cite{mcenery}. 
Each layer of the tracker is modelled as a %1 m by$  
1 m$^{2}$ sheet of pixelated silicon with 1 cm between layers. 
Since the reconstruction of events in a Compton telescope depend strongly on the energy resolution and position resolution, we were interested to explore the effect of the pixel size. We tested square pixels from 0.01 mm$^2$ up to 10.0 mm$^2$.
%The size of the squared pixels is the parameter we want to optimize and varies between 0.01 mm$^2$ up to 10.0 mm$^2$.
The thickness of the silicon layers is set to 500 $\mu$m for the results shown here %baseline geometry, 
but we have simulated additional thicknesses to understand the instrument response %have been tested 
(100 $\mu$m and 700 $\mu$m). %The distance between the tracker layers is set to 1 cm and the whole instrument is in void. 
We also studied the effect of %the 
passive material by applying additional non-sensitive silicon 
%layers 
material at 0.1 cm beneath each sensitive layer. 
The thickness of the passive layers varies so that the total amount of passive material represents a given percentage of the total active tracker mass. 
An
%The
ideal calorimeter is located underneath the tracker and is modelled as a 3-D position sensitive detector %very fine voxelled detector, 
whose function is to provide %the best reconstructed energy estimation.
an accurate measure of the Compton-scattered photon so that the calorimeter itself does not limit the telescope performance.
For every configuration, we simulate a monochromatic far-field point source %. We repeat the simulations for four energy values of the simulated source: 
at four energy values: 200~keV, 300~keV, 500~keV and 1~MeV.

Two of the main figures of merit for any imaging telescopes are the effective area and the angular resolution. 
The effective area is a measure of the efficiency of a telescope, and here we report results of the simulation as the number of events which were properly reconstructed and passed the event selections (see below) divided by the initial number of simulated events.
For Compton telescopes, the angular resolution is given by the FWHM of the distribution of the angular resolution measure (ARM) for each event, where the ARM is the smallest angular distance between the event circle and the true position of the source. %which evaluates effectiveness of the instrument and of the reconstruction algorithm performance in terms of reconstructed incoming photon direction. 

We adopt a classic Compton sequence event reconstruction algorithm available in MEGALib-revan library\footnote{http://megalibtoolkit.com/documentation.html}.
The ARM is evaluated on a subset of selected events whose reconstructed energies fall under the peak of the distribution centered on the true energy of the simulated source.

\begin{figure}
    \centering
    \includegraphics[scale=0.55]{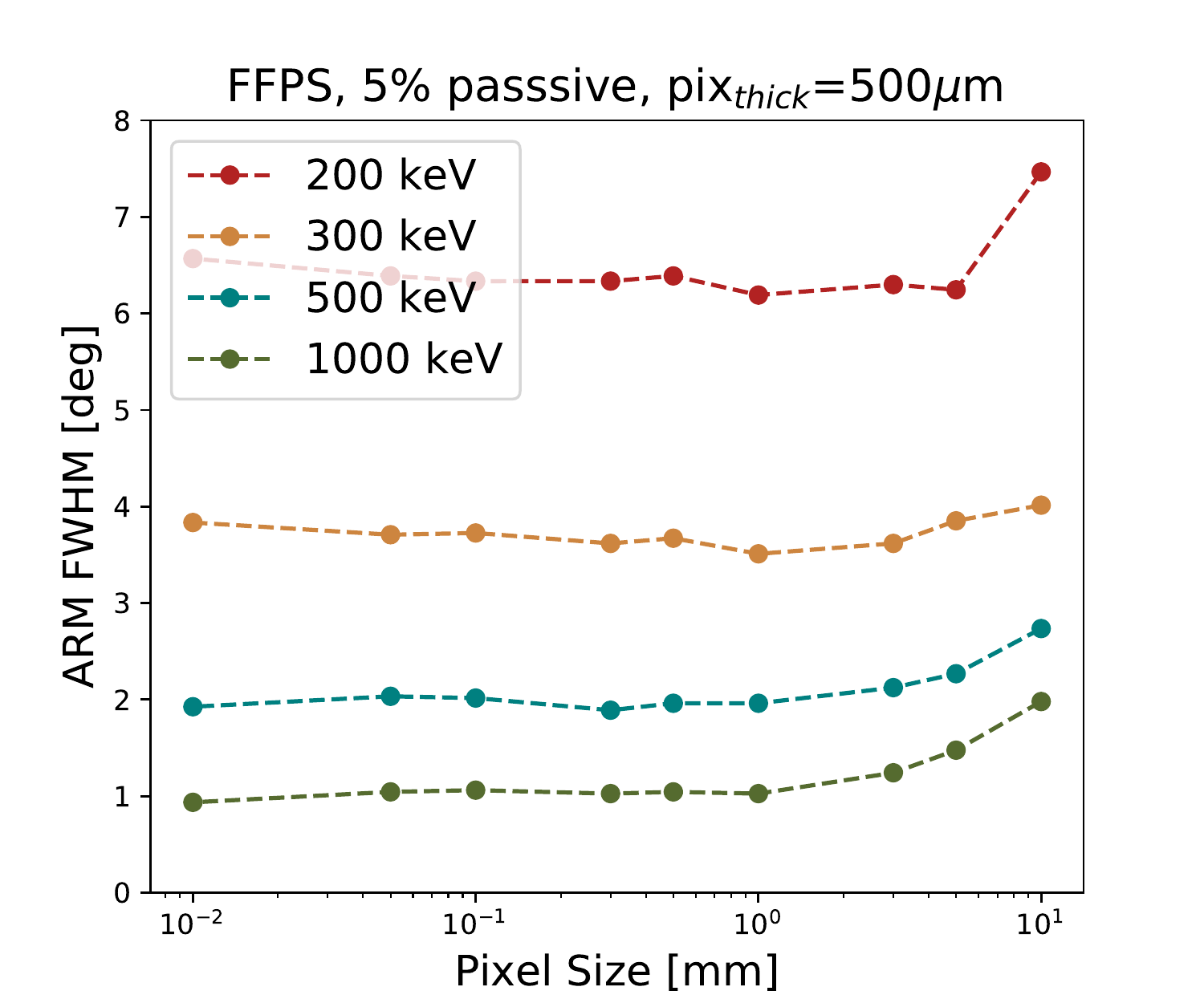}
    \caption{Angular resolution %ARM 
    as a function of the pixel size for different monochromatic energy simulated beams for 500 $\mu$m thick pixels. The fraction of passive material is set to 5$\%$.}
    \label{fig:ARMenergy}
\end{figure}

Figure \ref{fig:ARMenergy} illustrates the angular resolution %ARM 
as a function of the pixel size for different monochromatic sources for 500 $\mu$m thick pixels. 
We observe the expected trend of the improvement of the ARM with increasing energy. Also, for all the energies, we can appreciate a plateau in the ARM curve as a function of the pixel size up to $\sim$ 1 mm. This suggests that we could drastically enlarge the pixel size without loosing angular resolution, which results in a lower number of readout channels and consequently in a lower amount of total power supply (which is crucial aspect for any space-based instrument).

\begin{figure}
    \centering
    \includegraphics[scale=0.55]{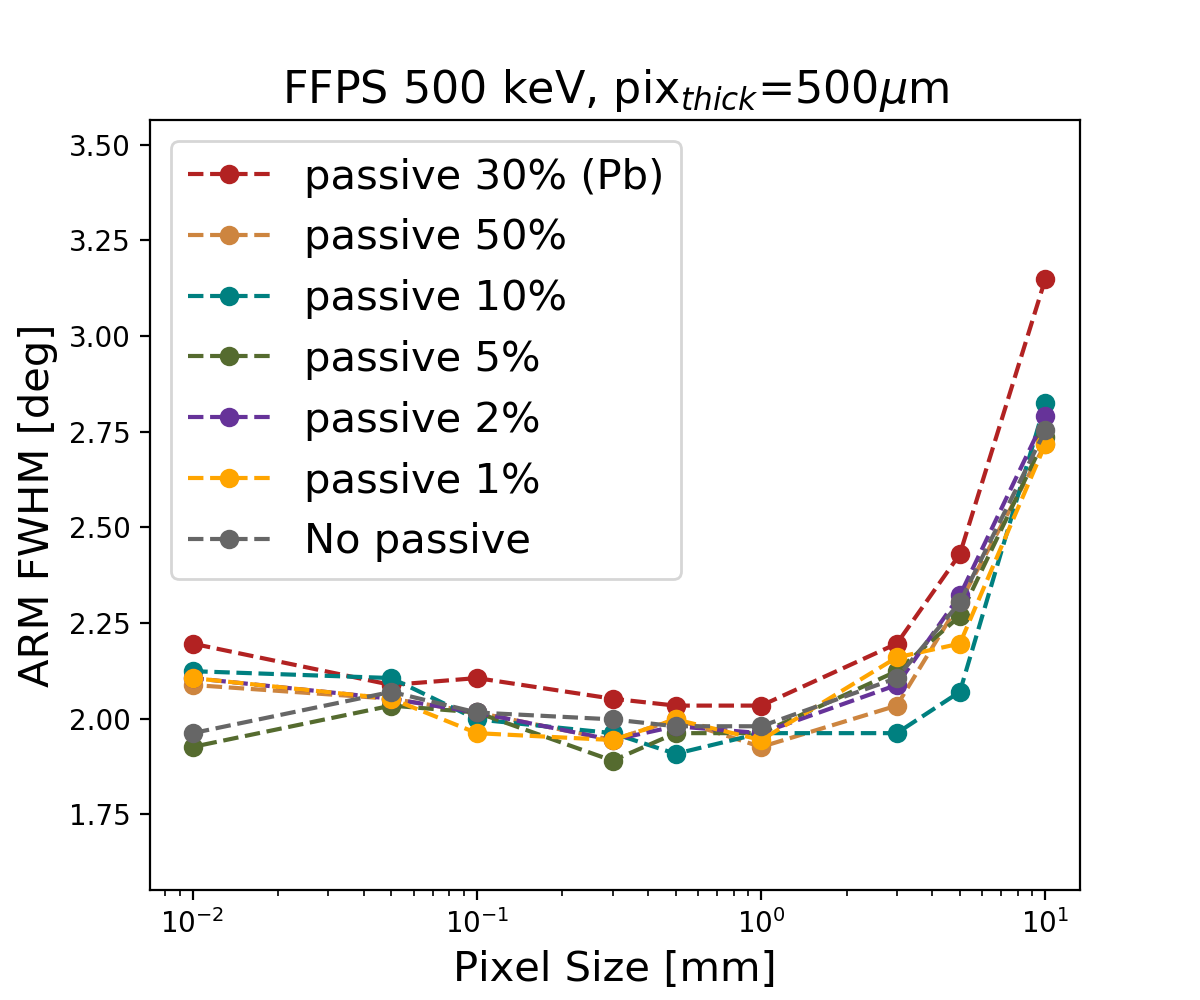}
    \includegraphics[scale=0.55]{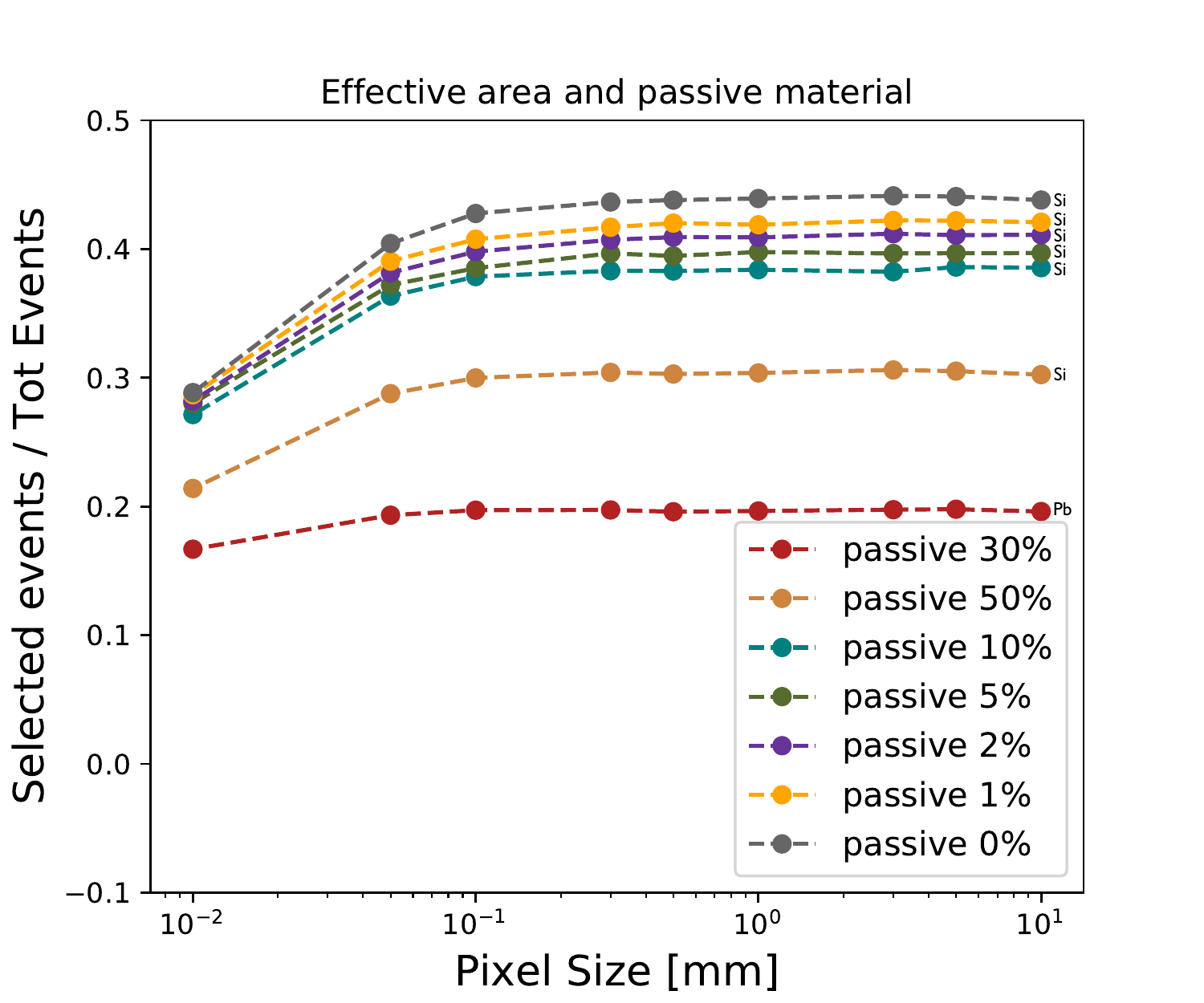}
    \caption{Study of the effect of the passive material in the tracker. Left: ARM as a function of the pixel size for different fractions (and kind) of passive material, Right: fraction of events passing the selection cuts as a function of the pixel size for different fractions (and kind) of passive material. Note that in both plots the 30$\%$ case uses lead as passive material.)}
    \label{fig:ARMpassive}
\end{figure}

Figure \ref{fig:ARMpassive} shows the effect of the passive material on the angular resolution %ARM 
and the fraction of events passing the energy cuts. For this study we focus on a median energy value (500 keV). As we can observe from the plot on the left, the amount of passive material does not affect the ARM estimation (only when we simulate 30\% of lead between the detector layers we start to appreciate a slight worsening of the ARM at every pixel sizes). The presence of the passive material, on the other hand, dramatically affects the efficiency of the event selection (plot on the right), and hence the statistics we can accumulate. Also, we observe that the efficiency starts to systematically decrease when the pixel size becomes lower that 0.1 mm: this is the effect of the smaller pixel volumes, which allow less energy to be released by the crossing particle, while the energy threshold to trigger the pixel is kept fixed to 20 keV in our simulation setup.

The results of the ARM and pixel size simulations led to the 500 $\mu$m pixel size and 500 $\mu$m Si depth requirements seen in Section \ref{sec:intro}. At 500 $\mu$m, the angular resolution of reconstructed Compton events is not limited by pixel size.


\section{CONCLUSION}
\label{sec:conclusion}

The results from ATLASPix were encouraging, and also reveal areas of improvement for future generations of the technology. We see that reasonable energy resolution is attainable with the detector, but currently only by extracting the analog output of individual pixels. This provides an opportunity to upgrade the digital output of future detectors by optimizing them for energy resolution instead of timing resolution. Other parameters, such as change in pixel size and depth, would make this technology increasingly optimized for the detection of high energy photons. A new version of the pixels called AstroPix V1 have been created and are currently undergoing testing. 

This is an exciting moment for future missions, with a new wave of telescopes on the horizon. A Si technology that economizes on payload space and has the potential to increase scientific parameters such as energy resolution and angular resolution would greatly assist with the next generation of gamma-sensitive missions. The eventual goal is to build a Compton telescope with AstroPix, which will give a better sense of the performance of this material in high-energy astrophysics.

\subsection{Outlook to AstroPix V1}

During the first quarter of 2020, the opportunity was given to submit a Pixel Matrix Sensor design using the 180 nm HV process from TSI Semiconductor. The design was implemented by Prof. Ivan Peric's group at Karlsruhe Institute of Technology, and aims at testing a pixel design heading towards the desired specification outlined through this paper: high dynamic range and low power consumption, since the primary application targets are orbital or deep-space flights.

\begin{figure}[h]
    \centering
    \includegraphics[scale=0.40]{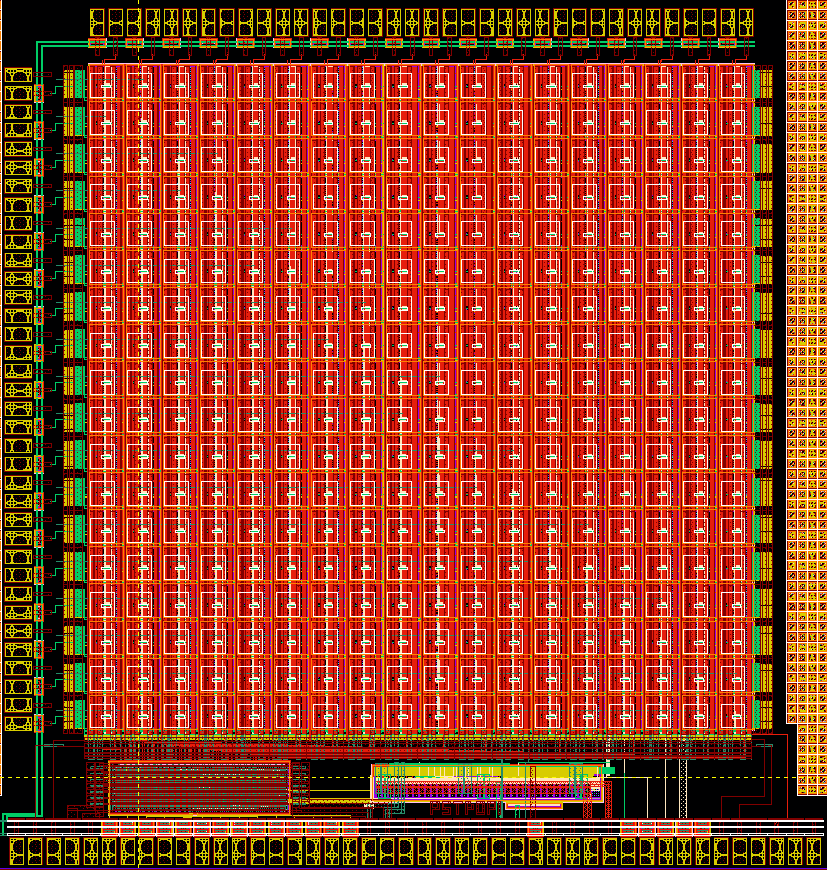}
    \caption{Top level view of the AstroPix V1 sensor matrix. 18 by 18 200 $\mu$m pixel matrix in the center, with digital readout and current biasing in the bottom peripheral area.}
    \label{fig:AstropixTopv1}
\end{figure}

The sensor is an 18 by 18 pixel matrix, with a pixel size of 200 by 200 $\mu$m, each pixel including an NMOS amplifier and PMOS comparator.  By increasing the pixel size, we are reducing the number of active circuits per cm$^2$, hence the total power consumption. 
The signal in-pixel comparator also features an additional custom Deep P implant layer, which isolates PMOS transistors from the sensitive diode Deep N implant, so that crosstalk between signal and  electronics can be reduced. This reduced crosstalk allows lowering the amplifier and comparator bias currents, hence the power consumption. 

For characterisation of this new pixel type, the sensor readout was greatly simplified by directly routing 36 pixel analog outputs to external pads, and by implementing the time over threshold digitisation as a simple 12 bit counter driven by an external clock. 

Initial measurements will begin in December 2020, and should lead to an AstroPix V2 revision which could be used for a first experimental orbital flight.

% References

\section{REFERENCES}

\hfill\begin{minipage}{\textwidth}

[1]  McEnery, J., Barrio, J. A., Agudo, I., and et al., \enquote{All-sky Medium Energy Gamma-ray Observatory:  Exploring the Extreme Multimessenger Universe,} Bulletin of the American Astronomical Society Vol.51, Issue7, 285–320, 363–376 (2019).

[2]  Caputo,  R. and et al.,  \enquote{Astropix:  Developing Silicon Pixel Detectors for Gamma-ray and Cosmic-ray Astrophysics,} APRA Proposal: Submitted in response to NNH18ZDA001N-APRA: D. 3 Astrophysics Research and Analysis, (2018).
 
[3]  Griffin,  S.,  \enquote{Development  of  A  Silicon  Tracker  for  the  All-sky  Medium  Energy  Gamma-ray  Observatory  Prototype,} in [AAS/High Energy Astrophysics Division], AAS/High Energy Astrophysics Division 17, 109.01 (Mar 2019).

[4]  Schoning,  A.,  Anders,  J.,  Augustin,  H.,  Benoit,  M.,  Berger,  N.,  Dittmeier,  S.,  and  et  al.,  \enquote{MuPix  and ATLASPix – Architectures and Results,} arXiv(2020).

[5]  H., B., \enquote{Zur Theorie des Durchgangs schneller Korpuskularstrahlen durch Materie,} Physik5, 325–400 (1930).

[6]  H., B., \enquote{Zur Theorie des Durchgangs schneller Elektronen durch Materie,} Physik14, 531–585 (1932).

[7]  F.,  B.,  \enquote{Zur  Bremsung  rasch  bewegter  Teilchen  beim  Durchgang  durch  Materie,} Physikvols. 16, 81, 285–320, 363–376 (1933).

[8]  Peric, I., Prathapan, M., Zhang, H., Weber, A., and Messaoud, F. G., \enquote{Description of the ATLASPixSimple and ATLASPixm2, preliminary v2.}

[9]  Zoglauer, A., Andritschke, R., and Schopper, F., \enquote{MEGAlib: The Medium Energy Gamma-ray Astronomy Library,} 50, 629–632 (Oct 2006).

\end{minipage}

\end{document}